\documentclass[runningheads]{llncs}
\usepackage{graphicx}
\usepackage{soul,color}
\usepackage{subcaption}
\usepackage{hyperref}
\usepackage{verbatim}

\begin{document}
\title{Constraint-Aware Trajectory for Drone Delivery Services\vspace{-0.4 cm}}
\authorrunning{J. Janszen et al.}
\author{
Jermaine Janszen \and
Babar Shahzaad \and
Balsam Alkouz \and
Athman Bouguettaya}


\institute{University of Sydney, Australia \\
\email{jjan3640@uni.sydney.edu.au}
\email{\{babar.shahzaad,balsam.alkouz,athman.bouguettaya\}@sydney.edu.au}}

\maketitle              % typeset the header of the contribution
\vspace{-0.5 cm}
\begin{abstract}
Drones are becoming a novel means for delivery services. We present a demonstration of drone delivery services in a skyway network that uses the service paradigm. A set of experiments is conducted using Crazyflie drones to collect the data on various positions of drones, wind speed, wind direction, and battery consumption. We run the experiments for a range of flight patterns including linear, rectangular, and triangular shapes. Demo: \url{https://youtu.be/tlXnUSIrRp0}
\vspace{-0.2 cm}
\vspace{-0.3 cm}
\end{abstract}

\section{Introduction}
\vspace{-0.2 cm}
Drones are autonomous aircraft that offer potential benefits for a multitude of civilian applications \cite{8682048}. Drones enable new services in various domains such as surveillance, agriculture, and delivery of goods \cite{10.1145/3460418.3479289}. Companies such as Amazon and Google are investing in the use of drones for delivery services \cite{9284115}. The targeted \emph{beneficiaries} of drone delivery services include consumers, transport companies, and suppliers of goods (e.g., medical suppliers, retailers, etc) \cite{shahzaad2021robust}.

Current research focuses on developing techniques for fast and cost-efficient deliveries using drones \cite{10.1007/978-3-030-33702-5_28}. However, existing works rely on simulation analysis of the proposed techniques using synthetic datasets. There is a paucity of real datasets especially those that include intrinsic and extrinsic factors affecting the drone services \cite{alkouz2020swarm}. Examples of intrinsic factors include flight range, battery, and payload capacity of a drone \cite{DBLP:journals/corr/abs-2107-05173}. Examples of extrinsic factors include number of recharging stations and weather conditions (e.g., wind speed and direction).

We leverage the \textit{service paradigm} to abstract a drone's capabilities as \textit{drone services}. Drone services usually operate in a skyway network taking into account no-flight and restricted zones \cite{SHAHZAAD2021335}. A skyway network is a set of predefined \textit{line-of-sight skyway segments}. A skyway segment is a straight line between two particular nodes. \textit{Each segment represents a service that is served by a drone.}

This demonstration focuses on drone delivery services in a skyway network to collect a real dataset that records the impact of different payloads and wind conditions over a set of trajectory patterns. We measure the battery consumption of a drone while carrying different payloads under varying wind conditions. We autonomously fly a drone through various flight patterns including \textit{triangular, rectangular, linear,} and \textit{hovering}. The collected dataset includes the drone’s XY positions, altitude, battery consumption, and wind for each flight. Finally, we plot the data collected on each flight to visualize and assess the impact of varying payloads and wind conditions on the battery consumption rate of the drone.

\begin{figure*}[t]
\minipage{0.32\textwidth}
\centering
  \includegraphics[width=\linewidth]{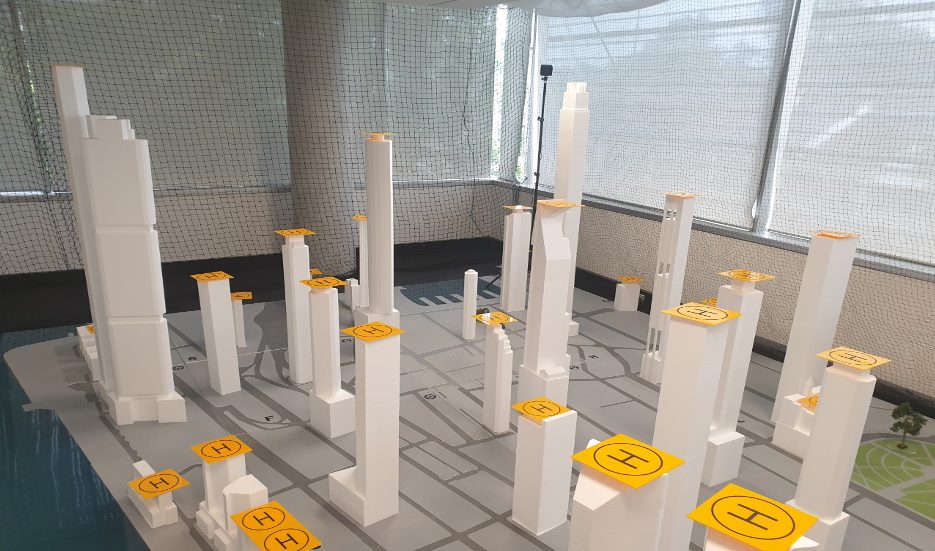}
   \caption{3D Model of Sydney CBD}\label{3d}
\endminipage\hfill
\minipage{0.32\textwidth}%
\centering
  \includegraphics[width=0.92\linewidth]{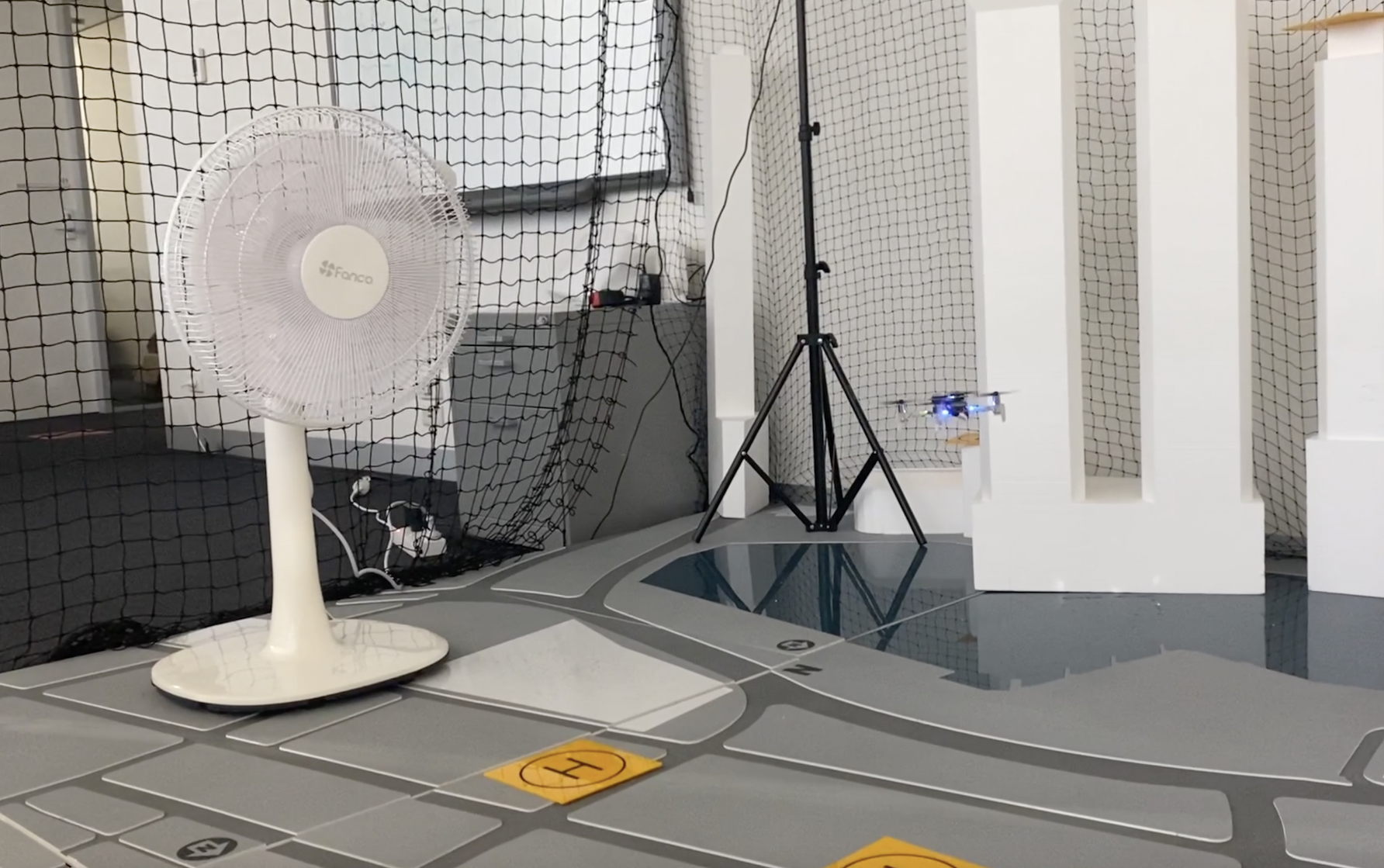}
   \caption{Fan for Wind Speed Control}\label{fan}
\endminipage\hfill
\minipage{0.32\textwidth}
  \includegraphics[width=0.93\linewidth]{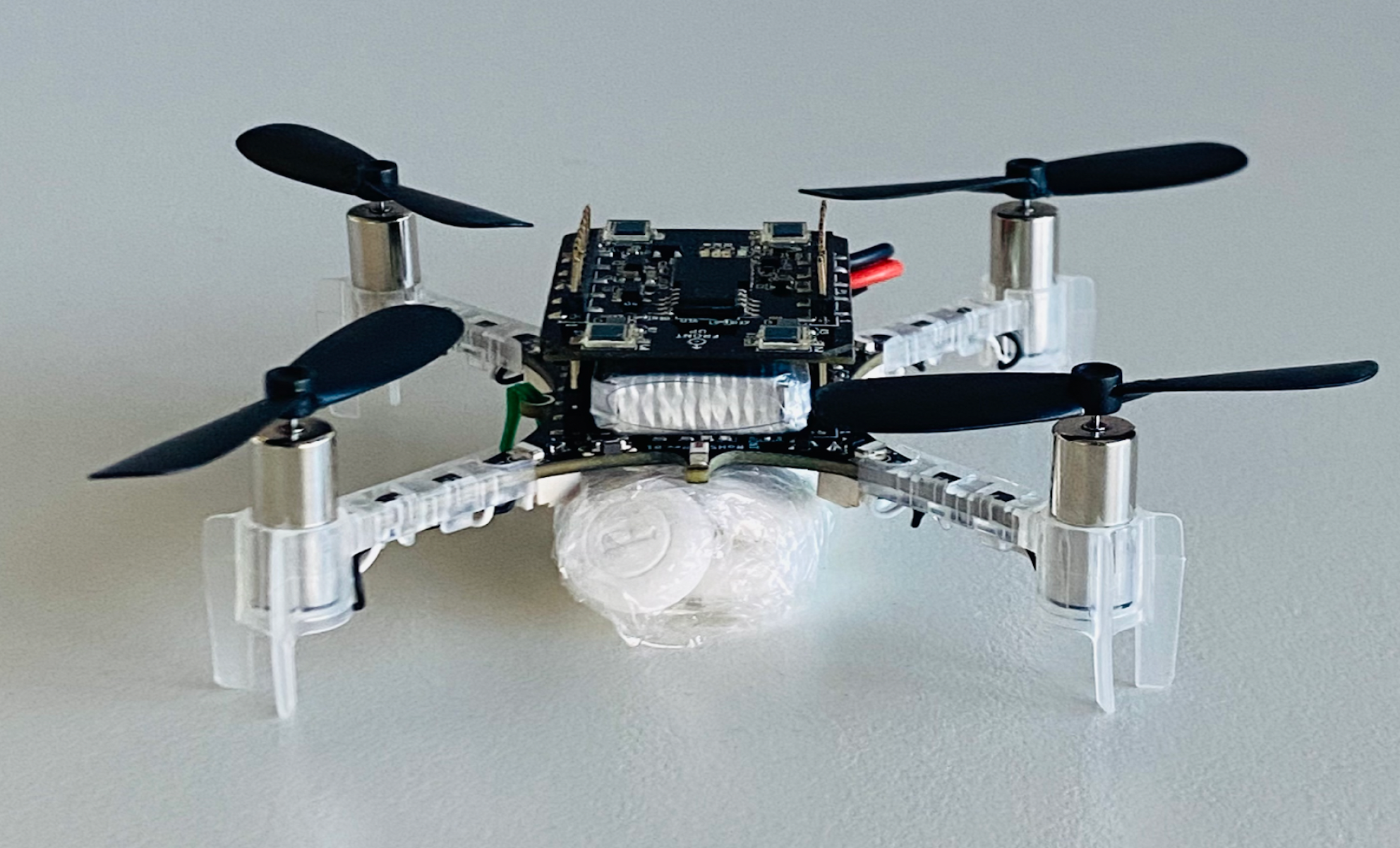}
  \caption{Crazyflie Carrying Payload}\label{payload}
\endminipage
\vspace{-0.6 cm}
\end{figure*}
\vspace{-0.4 cm}
\section{Trajectory Tracking and Data Collection}
\vspace{-0.3 cm}
We gather data on various drone parameters under the considered intrinsic and extrinsic factors with a \textit{focus on measuring battery consumption rate}. The experiments were performed to empirically measure the energy use of small drones. To run the experiment, the Crazyflie 2.1 drone by Bitcraze was used as it provides a modular setup with a python API. We setup a 3D model of Sydney CBD as an indoor testbed to mimic a skyway network (Fig. \ref{3d}). The drone locates its precise position during its flight with the aid of HTC Vive base stations, fitted at the corners of the lab. A fan with different speed settings is used to simulate extrinsic constraints (Fig. \ref{fan}). The drone is fitted with a payload to simulate intrinsic constraints (Fig. \ref{payload}).
Two main sets of trajectories were collected including hovering flights and predefined flight paths trajectories.
\vspace{-0.4 cm}
\subsection{Hovering Flight with Different Conditions}
\vspace{-0.1 cm}
\subsubsection{Hovering with Extended Flight Time.}
This test measures how the battery voltage changes over time. The drone hovered in a fixed position 50cm off the ground with no payload and no wind. The drone maintained its position for 5 minutes as various drone parameters were logged in 100ms intervals.
\vspace{-0.3 cm}
\subsubsection{Hovering with Different Wind Speeds.}
This test measures how the battery voltage changes over time under different wind speeds. The drone hovered in a fixed position 60cm off the ground and 3m away from a fan with the same height. The drone would hover for 2 minutes while parameters were logged in 50ms intervals. To procure a robust set of data, 4 different wind speeds (1.8 km/h, 2.2 km/h, 2.9 km/h, and 3.6 km/h) were used, with each speed being tested twice. Greater speeds weren't used as the drone's stability was greatly compromised above 4 km/h when hovering.
\vspace{-0.4 cm}
\subsubsection{Hovering with Different Payload Weights.}
This test measures how the battery voltage changes over time with varying payload weights. The drone hovered in a fixed position 50cm off the ground while carrying different payload weights. The drone would hover for 2 minutes while parameters were logged in 50ms intervals. Four different payload weights (2g, 4g, 6g, and 8g) were used to procure a robust set of data, with each payload being tested 4 times.
\begin{figure}[t]
  \begin{subfigure}[t]{0.32\textwidth}
    \includegraphics[width=\linewidth]{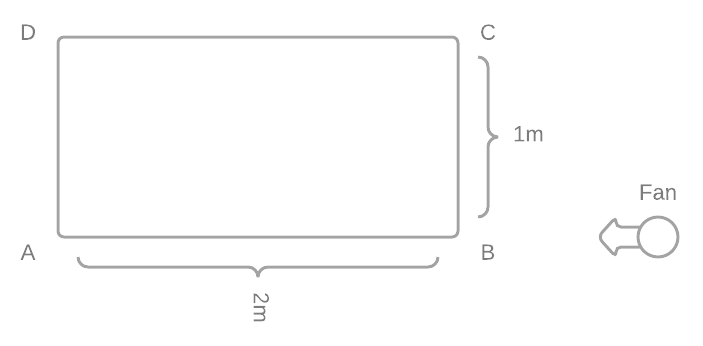}
    \vspace{-0.4 cm}
   \caption{Rectangular Flight Path}\label{rectangle-diagram}
  \end{subfigure}\hfill
  \begin{subfigure}[t]{0.32\textwidth}
    \includegraphics[width=0.8\linewidth]{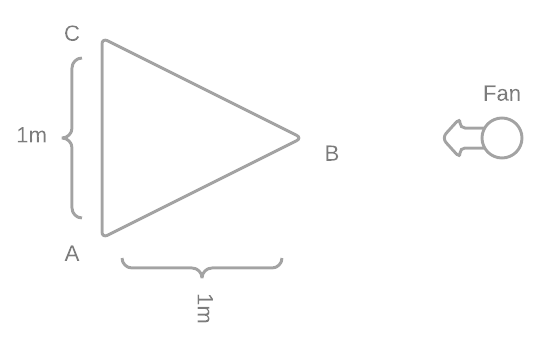}
    
   \caption{Triangular Flight Path}\label{triangle-diagram}
  \end{subfigure}
  \begin{subfigure}[t]{0.32\textwidth}
    \includegraphics[width=\linewidth]{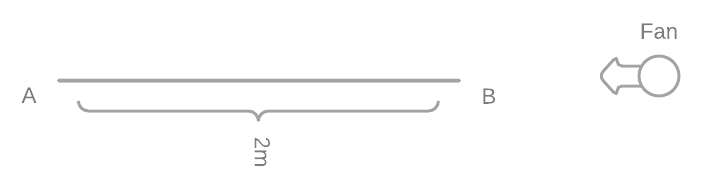}
    \vspace{-0.4 cm}
    \caption{Linear Flight Path}\label{line-diagram}
  \end{subfigure}
    \vspace{-0.2 cm}
  \caption{Flight Paths with Fan Placement and Direction}
  \label{fig:paths}
\vspace{-0.3 cm}
\end{figure}
\vspace{-0.2 cm}
\subsection{Fixed Flight Paths with Different Conditions}
\vspace{-0.1 cm}
In addition to the hovering tests, the drone parameters were also logged during the drone flights over predefined paths. The selected paths include a rectangle, triangle, and line which allow the drone to fly forwards, backwards, side to side, and diagonally. These directions are the typical directions that drones follow traversing the nodes in a skyway network. Fig.\ref{fig:paths} illustrates different flight paths collected in the dataset. Throughout the flights, the drone travelled at a fixed speed of 0.3 m/s at 0.5m height and hovered for 5 seconds at each endpoint along the path. To generate control data points with drone only settings, the paths were flown without any payload or wind. We again run the tests with the same payloads as used for the hovering tests. We used a fan to generate artificial wind. For the rectangle and line paths, the fan was placed 1m away from point B facing back at point A while the fan was placed 1m away from point B facing the midpoint between points A and C for the triangle path.
\vspace{-0.4 cm}
\section{Data}
\vspace{-0.2 cm}
We performed 72 flights under a number of operational parameters (payloads, wind speeds). In addition, 34 recordings were performed to assess the drone’s ancillary power and hover conditions. The data collected from the experiments are organized in CSV files where each sheet contains a log under different settings.

\begin{figure*}[t]

\minipage{0.32\textwidth}
\centering
  \includegraphics[width=0.92\linewidth]{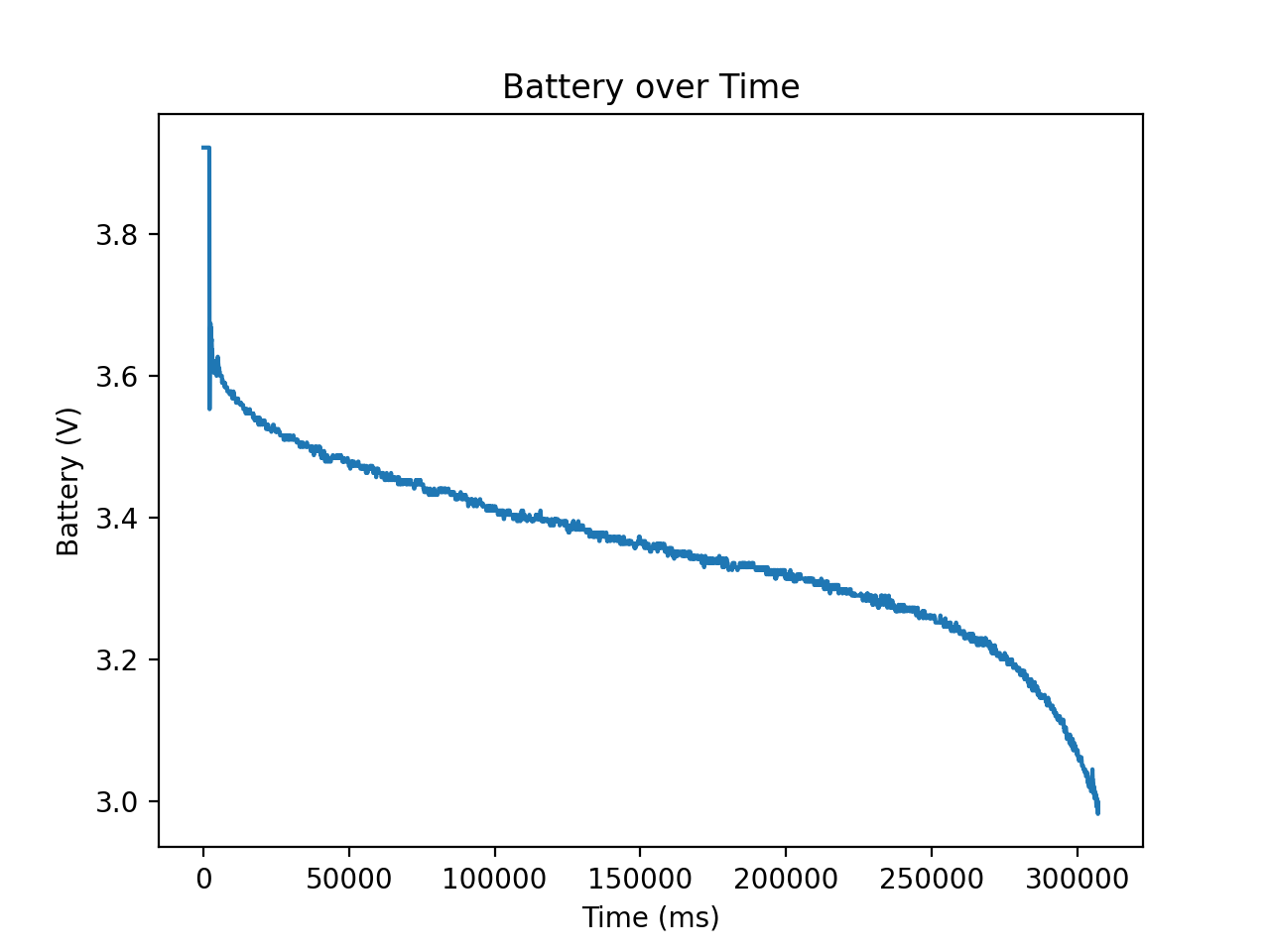}

   \caption{Battery Consumption When Hovering}\label{hover-battery}
\endminipage\hfill
\minipage{0.32\textwidth}%
\centering
  \includegraphics[width=0.92\linewidth]{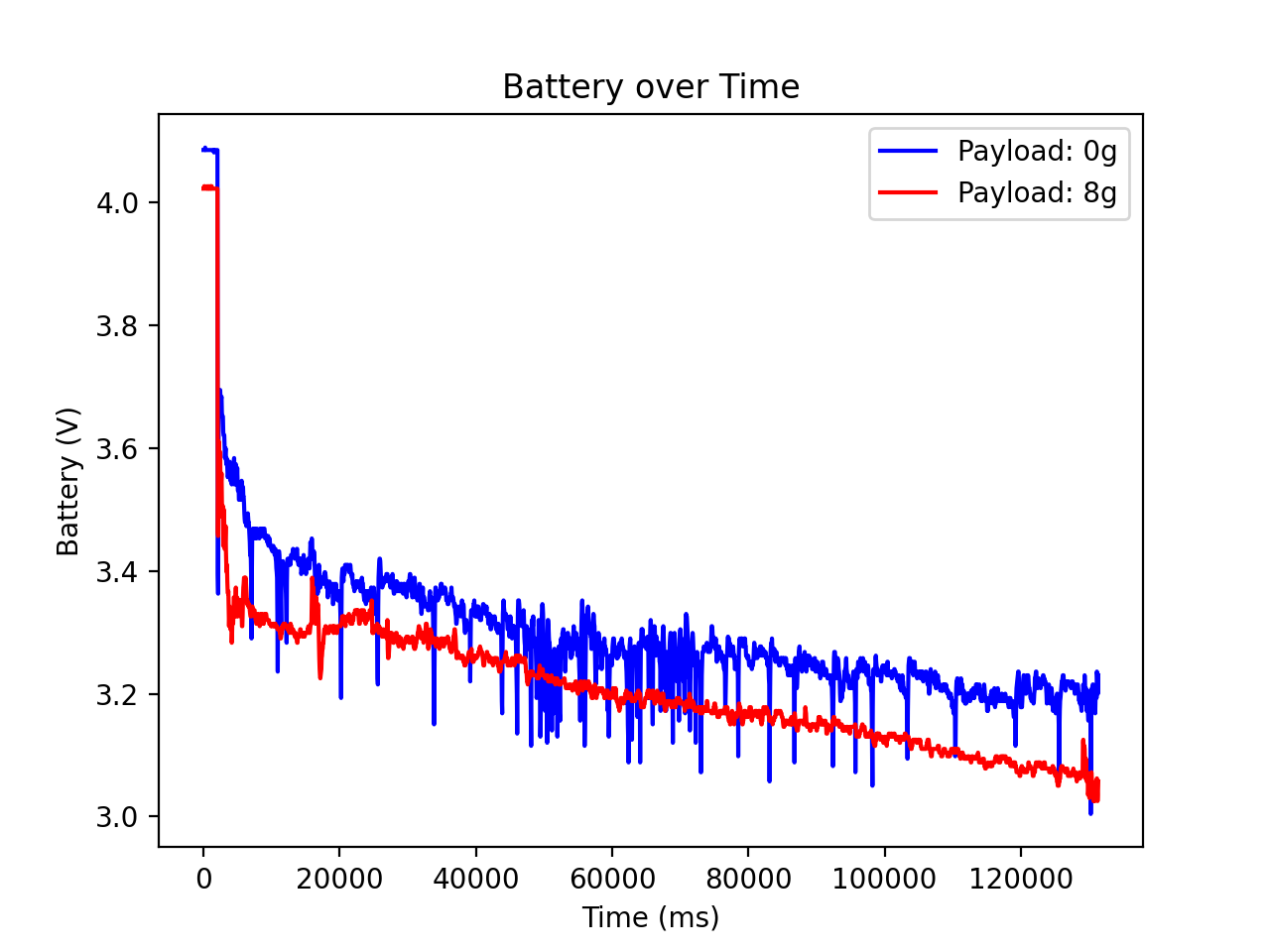}

   \caption{Battery Consumption in Hovering State}\label{hover-battery-payload}
\endminipage\hfill
\minipage{0.32\textwidth}
  \includegraphics[width=0.92\linewidth]{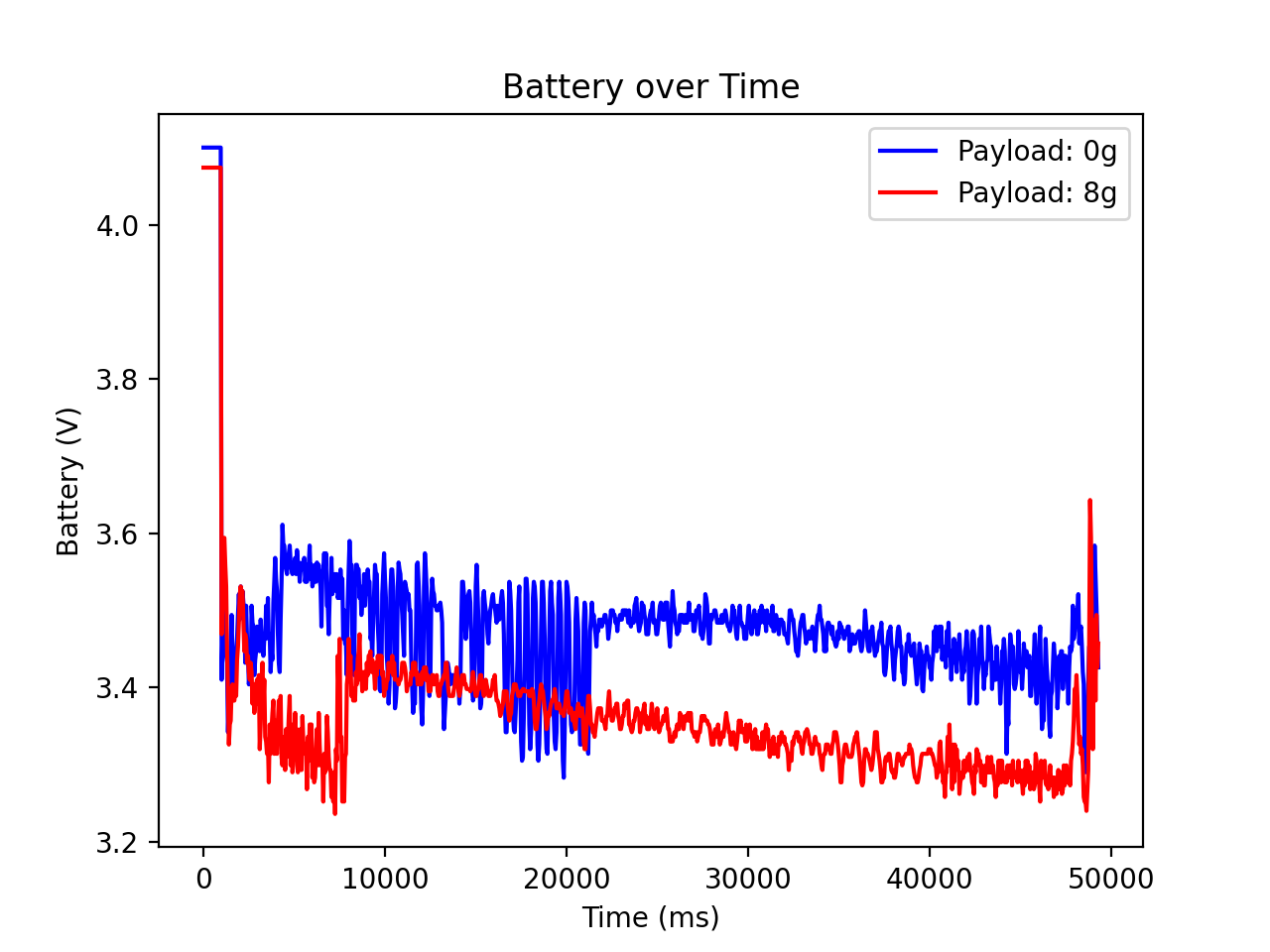}

  \caption{Battery Consumption in Rectangular flight}\label{rectangle-battery-payload}
\endminipage
\vspace{-0.4 cm}
\end{figure*}

\begin{figure*}[t]

\minipage{0.32\textwidth}
\centering
  \includegraphics[width=0.92\linewidth]{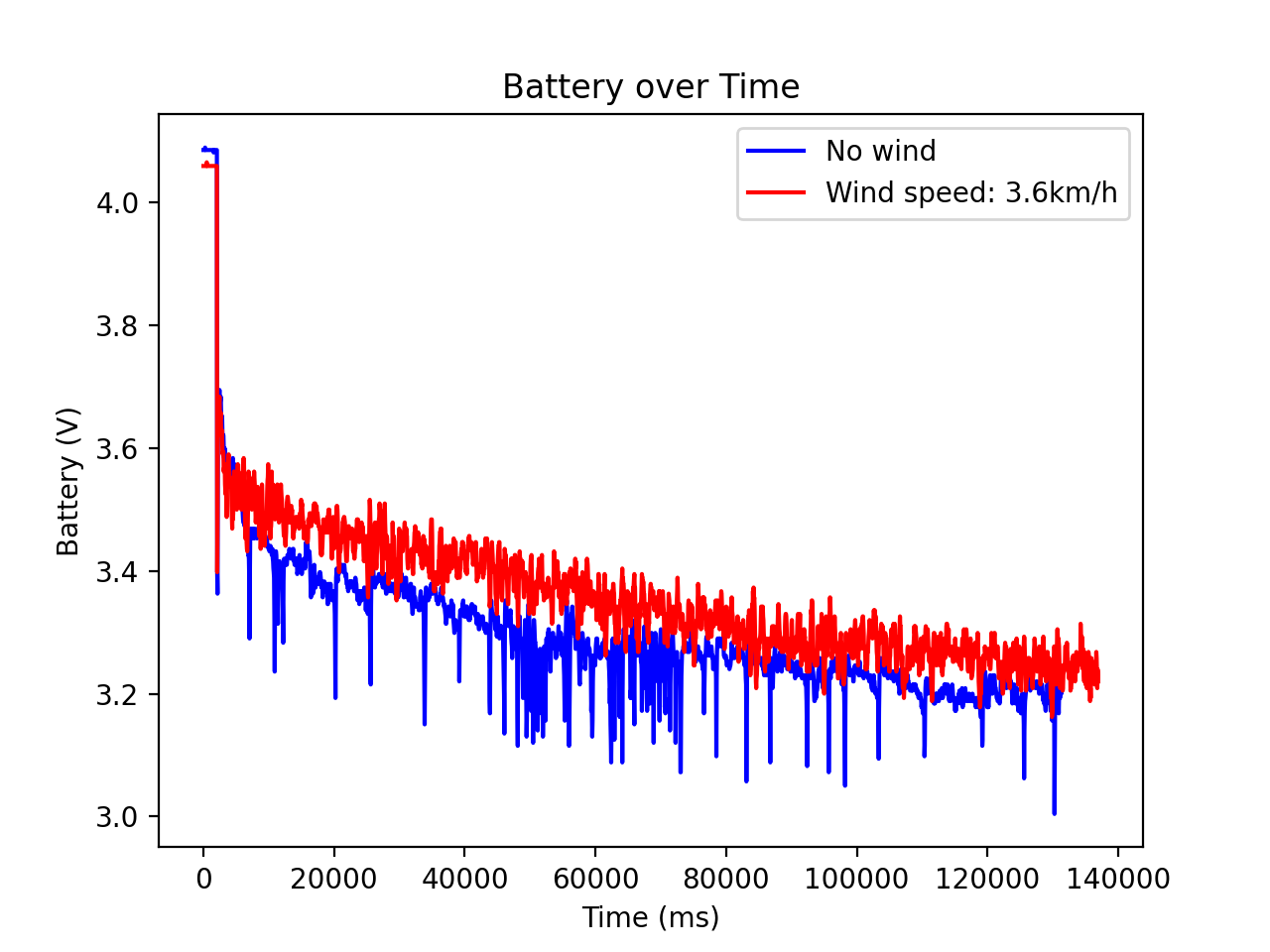}

\caption{Battery Consumption in Hovering State} \label{hover-battery-wind}
\endminipage\hfill
\minipage{0.32\textwidth}%
\centering
  \includegraphics[width=0.92\linewidth]{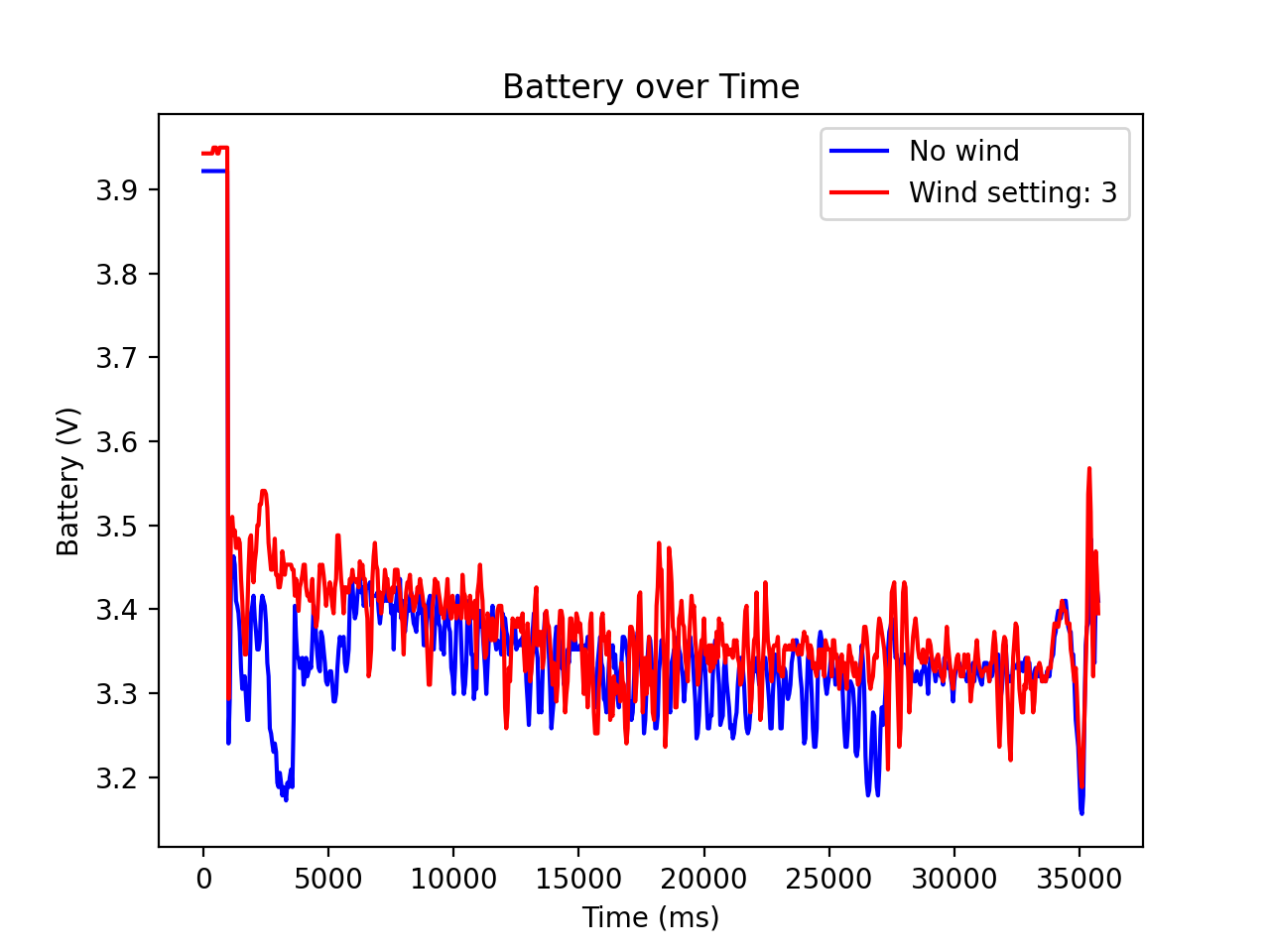}

\caption{Battery Consumption in Triangular} \label{triangle-battery-wind}
\endminipage\hfill
\minipage{0.32\textwidth}
  \includegraphics[width=0.92\linewidth]{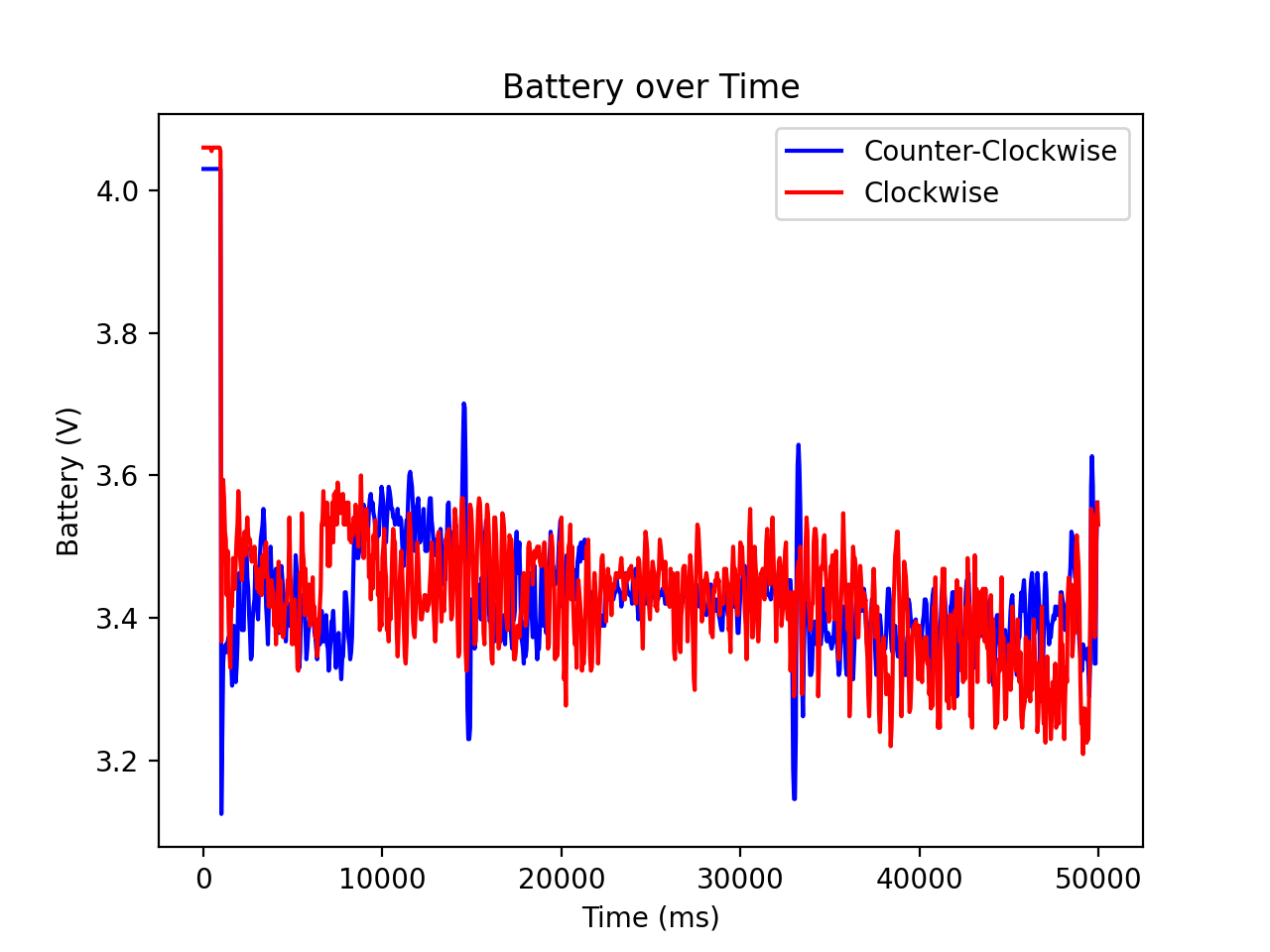}

\caption{Battery Consumption in Rectangular} \label{rectangle-battery-counter}
\endminipage
\vspace{-0.4 cm}
\end{figure*}

\begin{figure*}[t]

\minipage{0.32\textwidth}
\centering
  \includegraphics[width=0.92\linewidth]{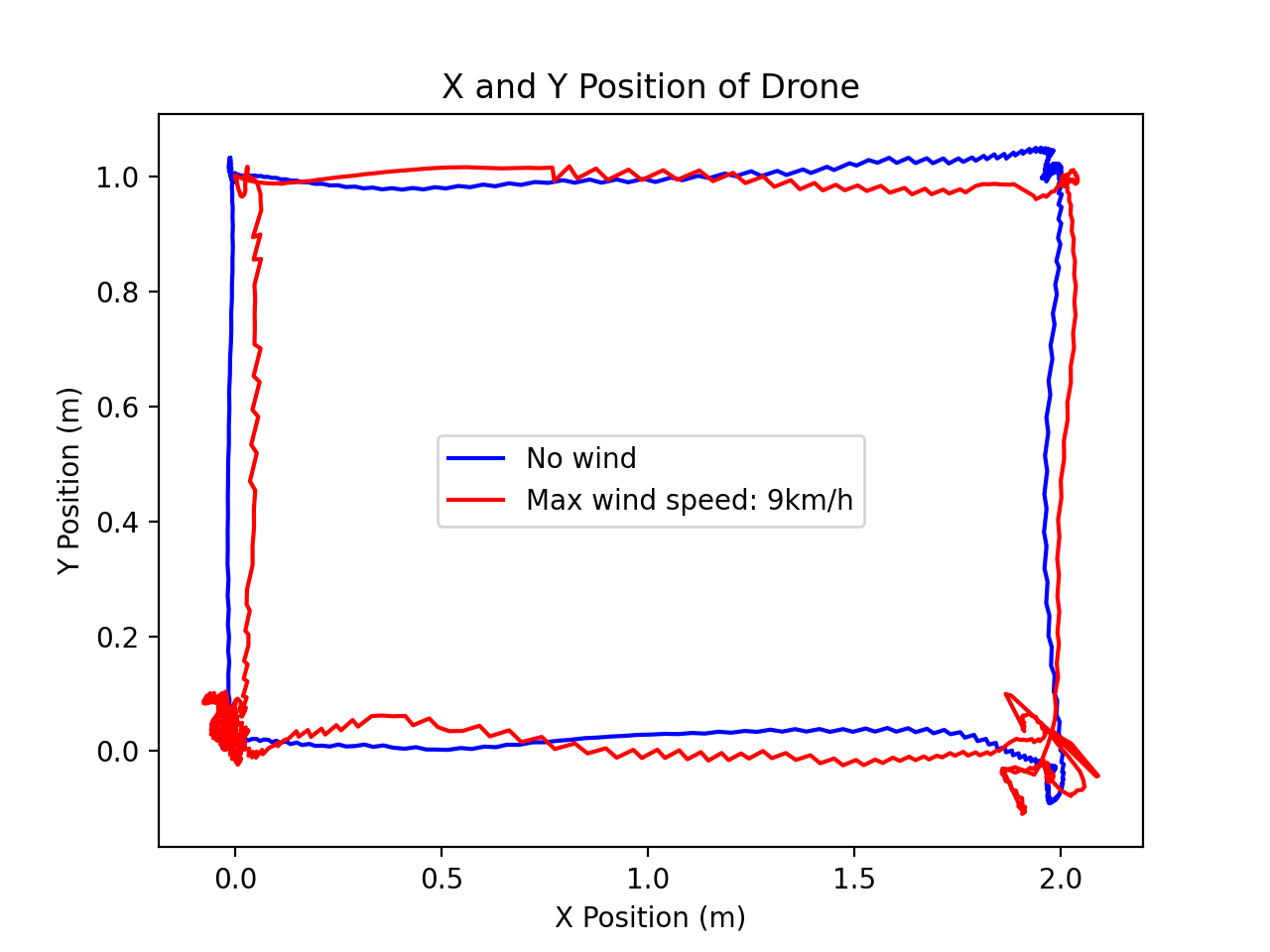}

\caption{Stability in Rectangular Path} \label{rectangle-position-wind}
\endminipage\hfill
\minipage{0.32\textwidth}%
\centering
  \includegraphics[width=0.92\linewidth]{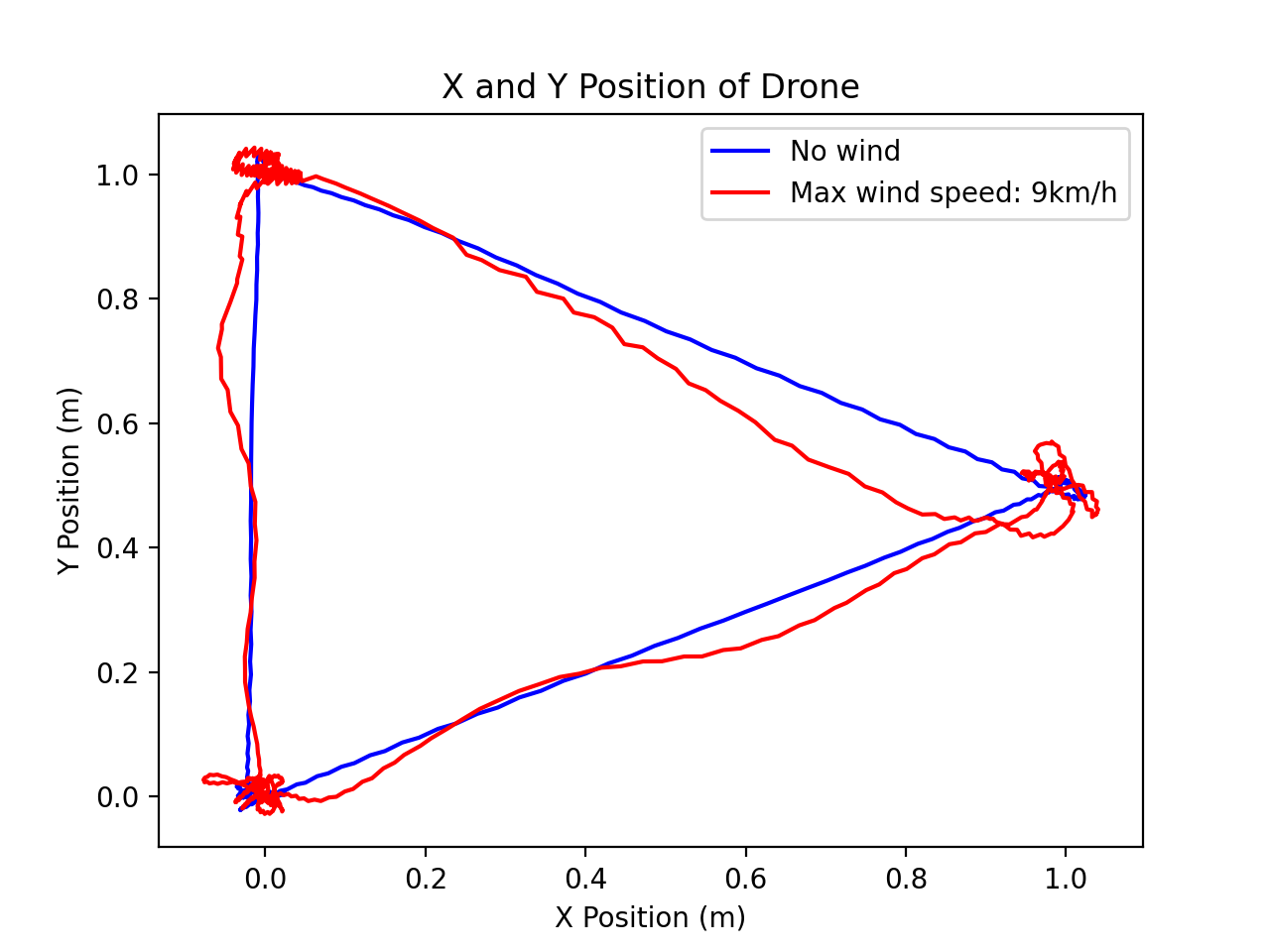}

\caption{Stability in Triangular Path} \label{triangle-position-wind}
\endminipage\hfill
\minipage{0.32\textwidth}
  \includegraphics[width=0.92\linewidth]{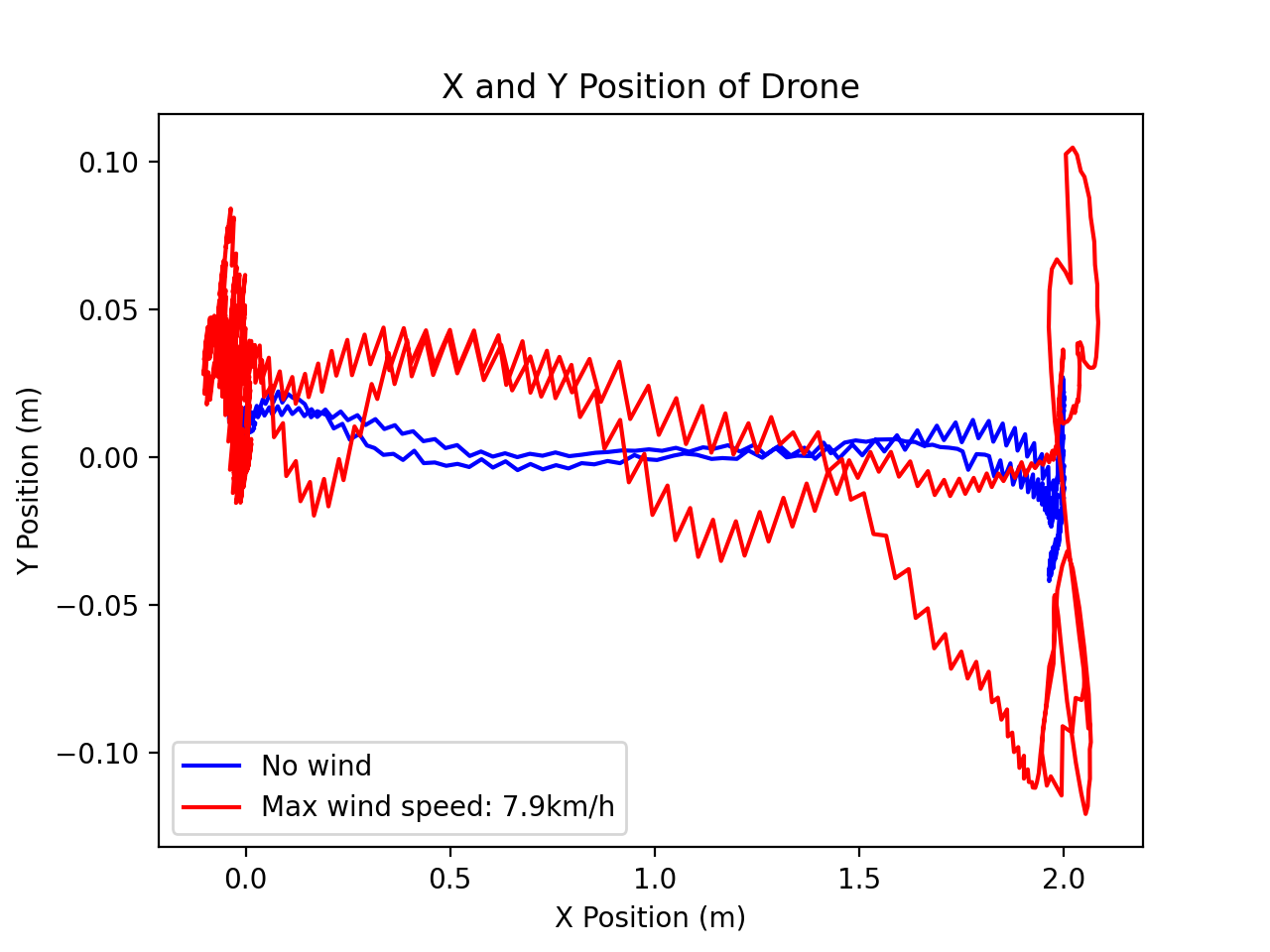}

\caption{Stability in Linear Path} \label{line-position-wind}
\endminipage
\vspace{-0.6 cm}
\end{figure*}

\vspace{-0.4 cm}
\section{Results}
\vspace{-0.2 cm}
In what follows, we present an analysis of the collected dataset in terms of battery consumption with different payloads under various wind conditions. Fig. \ref{hover-battery} shows the battery consumption trend over time without payload in the hovering state until the battery was drained. We observe that the battery consumption of a drone increases as the payload weight increases (Fig. \ref{hover-battery-payload} and \ref{rectangle-battery-payload}).
In addition, we observe that the battery consumption behaviour shows a fluctuating behaviour under different wind conditions (Fig. \ref{hover-battery-wind} and \ref{triangle-battery-wind}). This fluctuation is due to the changing wind directions during the drone flight. Flying a drone with a headwind is more energy-efficient \cite{alyassi2021autonomous}. Therefore, when the drone flies in a headwind direction, it consumes less battery compared to the tailwind direction. Fig. \ref{rectangle-battery-counter} presents the battery consumption behaviour while flying in a rectangular path in clockwise and anti-clockwise directions. Fig. \ref{rectangle-position-wind}, \ref{triangle-position-wind}, and \ref{line-position-wind} shows the stability of the drone under various wind conditions in different flight patterns. The drone's stability is highly affected when flying under strong wind conditions.
\vspace{-0.2 cm}
\subsubsection*{Acknowledgment.}
This research was partly made possible by DP160103595 and LE180100158 grants from the Australian Research Council. The statements made herein are solely the responsibility of the authors.

\vspace{-0.4 cm}
\bibliographystyle{splncs04}

\bibliography{references}
\vspace{-0.4 cm}
\end{document}